\documentclass[3p,times]{elsarticle}

\usepackage{ecrc}

\usepackage{epstopdf}

\volume{00}

\firstpage{1}

\journalname{Nuclear Physics A}

\runauth{U. Heinz et al.}

\jid{nupha}

\jnltitlelogo{Nuclear Physics A}

\usepackage{graphicx}
\usepackage{amsmath,amssymb}
\usepackage{bm}
\newcommand{\eq}{{\,=\,}}

\begin{document}

\begin{frontmatter}



\title{Viscous hydrodynamics for strongly anisotropic expansion}

%% Single author (and collaboration) - please insert
%\author{Author (for the XYZ\fnref{col1} Collaboration)}
%\fntext[col1] {A list of members of the XYZ Collaboration and acknowledgements can be found at the end of this issue.}
%\address{Address}

%% For multiple authors, replace the above by:

\author[OSU]{Ulrich Heinz}
\author[OSU]{Dennis Bazow}
\author[Kent]{Michael Strickland}

\address[OSU]{Department of Physics, The Ohio State University, Columbus, Ohio 43210-1117, USA}
\address[Kent]{Department of Physics, Kent State University, Kent, Ohio 44242, USA}

\begin{abstract}
A new formulation of second-order viscous hydrodynamics, based on an expansion around a locally
anisotropic momentum distribution, is presented. It generalizes the previously developed formalism of
anisotropic hydrodynamics ({\sc aHydro}) to include a complete set of dissipative currents for which equations of motion are derived by solving the Boltzmann equation in the 14-moment approximation. By solving the {\sc vaHydro} equations for a transversally homogeneous, longitudinally boost-invariant 
system ((0+1)-dimensional expansion) and comparing with the exact solution of the Boltzmann equation in relaxation-time approximation we show that {\sc vaHydro} performs much better than all other known second-order viscous hydrodynamic approximations.
\end{abstract}


\end{frontmatter}

% \linenumbers

\section{Introduction}
\label{intro}
%%%%%%%%%%%%%%%%%%%%%%%%%%%%%%%%%%%%%%%%%%%%%%%%%%%%%%%%

In relativistic heavy-ion collisions large differences between the longitudinal and transverse expansion rates lead to large shear viscous effects, generating large anisotropies between the longitudinal and transverse pressures during the early stage of their evolution. These cause standard Israel-Stewart (second-order) viscous hydrodynamic theory (see \cite{Gale:2013da} for a recent review) to break down. Anisotropic hydrodynamics \cite{Martinez:2010sc,Florkowski:2010cf,Tinti:2013vba} ({\sc aHydro}) deals with the large longitudinal/transverse pressure anisotropy, $\mathcal{P}_L{-}\mathcal{P}_T$, ``non-perturbatively'', thereby avoiding the occurrence of negative longitudinal pressures and improving the performance of hydrodynamics at early times \cite{Martinez:2010sc,Florkowski:2010cf}. However, {\sc aHydro} \cite{Martinez:2010sc,Florkowski:2010cf} accounts only for one (the largest) of the five independendent components of the shear stress tensor $\pi^{\mu\nu}$. It can therefore not be used to compute the viscous suppression of elliptic flow which is sensitive to e.g. $\pi^{xx}{-}\pi^{yy}$. On the other hand, since the four remaining components of the shear stress tensor never become as large as the longitudinal/transverse pressure difference (with smooth initial density profiles they start out as zero, and with fluctuating initial conditions they are initially small), they can be treated ``perturbatively'' \`a la Israel and Stewart, without running into problems even at early times. Combining the non-perturbative dynamics of $\mathcal{P}_L{-}\mathcal{P}_T$ via {\sc aHydro} with a perturbative treatment of the remaining viscous stress terms $\tilde{\pi}^{\mu\nu}$ \`a la Israel-Stewart defines our new {\sc vaHydro} scheme. It is expected to perform better than both IS theory and {\sc aHydro} during all evolution stages.

%%%%%%%%%%%%%%%%%%%%%%%%%%%%%%%%%%%%%%%%%%%%%%%%%%%%%%%
\vspace*{-3mm}
\section{Hydrodynamic approximations of kinetic theory}
\label{sec2}
\vspace*{-1mm}
%%%%%%%%%%%%%%%%%%%%%%%%%%%%%%%%%%%%%%%%%%%%%%%%%%%%%%%
The physics behind different hydrodynamic approximations to the underlying microscopic dynamics is best illustrated by starting from the Boltzmann equation and considering conditions (weakly coupled microscopic dynamics and small pressure gradients on the macroscopic level) where both approaches are simultaneously valid. The form of the resulting hydrodynamic equations (but not the value of the associated transport coefficients) is independent of the coupling strength and remains unchanged for a strongly coupled liquid. We restrict ourselves to conformal systems with massless degrees of freedom -- generalizations to massive systems can be found in the literature \cite{Florkowski:2014bba}.

{\bf 1.} Using the relaxation-time approximation (RTA) for the collision term, with $\tau_\mathrm{rel}(x)\eq{c}/T(x)$ where $T$ is the local temperature, the Boltzmann equation reads
\begin{equation}
\label{eq1}
p^\mu\partial_\mu f(x,p) = C(x,p) \equiv \frac{p{\cdot}u(x)}{\tau_\mathrm{rel}(x)}\,
    \Bigl[f_\mathrm{eq}(x,p){-}f(x,p)\Bigr].
\end{equation}
We define $p$-moments of the distribution function weighted with some momentum observable $O(p)$ by
\begin{equation}
\label{eq2}
 \langle O(p)\rangle \equiv \int_p O(p)\, f(x,p) \equiv \frac{g}{(2\pi)^3} \int \frac{d^3p}{E_p} O(p)\, f(x,p)
\end{equation}
($g$ is a degeneracy factor). The (baryon-)charge current and energy momentum tensor are then written as 
\begin{equation}
\label{eq3}
j^\mu\eq\langle p^\mu\rangle,\quad
T^{\mu\nu}\eq\langle p^\mu p^\nu\rangle.
\end{equation} 
They take their ideal fluid dynamical form $j^\mu_\mathrm{id}\eq{n}u^\mu$ and $T_\mathrm{id}^{\mu\nu}\eq{e} u^\mu u^\nu - \mathcal{P}\Delta^{\mu\nu}$ (where $\Delta^{\mu\nu}\eq{g}^{\mu\nu}{-}u^\mu u^\nu$ is the spatial projector in the local rest frame (LRF)) if we assume that the system is locally momentum isotropic:
\begin{equation}
\label{eq4}
f(x,p)=f_\mathrm{iso}(x,p)\equiv f_\mathrm{iso}\left(\frac{p{\cdot}u(x)-\mu(x)}{T(x)}\right).
\end{equation}
This ideal fluid decomposition does not require chemical equilibrium, nor does it require complete thermal equilibrium in the sense that the dependence of $f_\mathrm{iso}$ on its argument is exponential. If the dependence is non-exponential, the collision term in the Boltzmann equation is non-zero, but its $p^\mu$-moment still vanishes, $\int_p p^\mu C\eq0$, due to energy-momentum conservation. The ideal hydrodynamic equations follow by inserting this ideal fluid decomposition into
\begin{equation}
\label{eq5}
    \partial_\mu j^\mu = \frac{n_\mathrm{eq}-n(x)}{\tau_\mathrm{rel}(x)},\quad
    \partial_\mu T^{\mu\nu} = 0,
\end{equation}
which one can solve for the local charge density $n(x)$, energy density $e(x)$, and flow velocity $u^\mu(x)$, with temperature $T(x)$, chemical potential $\mu(x)$ and pressure $\mathcal{P}(x)$ following from the equation of state (EOS) of the fluid. Local deviations from chemical equilibrium result in a non-equilibrium value of the local chemical potential and a non-zero right hand side in the charge conservation equation. Deviations from thermal equilibrium (while preserving local isotropy) must be accounted for by a non-equilibrium pressure in the EOS\, $\mathcal{P}(e,n)$. In both cases Eqs.~(\ref{eq5}) lead to a non-vanishing entropy production rate $\partial_\mu S^\mu{\,\sim\,}1/\tau_\mathrm{rel}{\,\ne\,}0$.

{\bf 2.} Israel-Stewart (IS) second-order viscous fluid dynamics \cite{Israel:1979wp} is obtained by using in (\ref{eq3}) for $f(x,p)$ the ansatz
\begin{equation}
\label{eq6}
 f(x,p)= f_\mathrm{iso}\left(\frac{p{\cdot}u(x)-\mu(x)}{T(x)}\right) +\delta f(x,p).
\end{equation}
For later convenience we decompose $p^\mu$ into its temporal and spatial components in the LRF: $p^\mu\eq(u^\mu u^\nu{+}\Delta^{\mu\nu})p^\mu={E}u^\mu {+} p^{\langle\mu\rangle}$ where $E{\,\equiv\,}u{\cdot}p$ and $p^{\langle\mu\rangle}{\,\equiv\,}\Delta^{\mu\nu}p_\nu$. Then $n\eq\langle E\rangle$ and $e\eq\langle E^2\rangle$. The decomposition (\ref{eq6}) is made unique by Landau matching: First, define the LRF by solving the eigenvalue equation $T^{\mu\nu} u_\nu\eq{e}u^\mu$ with the constraint $u^\mu u_\mu\eq1$. This fixes the flow vector $u^\mu(x)$ and the LRF energy density. Next, we fix $T(x)$ and $\mu(x)$ by demanding that $\delta f$ gives no contribution to the local energy and baryon density: $\langle E\rangle_\delta = \langle E^2\rangle_\delta=0$. Inserting (\ref{eq6}) into (\ref{eq3}) we find the general decomposition
\begin{equation}
\label{eq7}
j^\mu = j^\mu_\mathrm{id}  + V^\mu, \quad
T^{\mu\nu}=T^{\mu\nu}_\mathrm{id} - \Pi\Delta^{\mu\nu} + \pi^{\mu\nu},
\end{equation}
with a non-zero charge flow $V^\mu\eq\bigl\langle p^{\langle\mu\rangle}\bigr\rangle_\delta$ in the LRF, a bulk viscous pressure $\Pi\eq-\frac{1}{3}\bigl\langle p^{\langle\alpha\rangle} p_{\langle\alpha\rangle}\bigr\rangle_\delta$, and a shear stress $\pi^{\mu\nu}\eq\bigl\langle p^{\langle\mu} p^{\nu\rangle} \bigr\rangle_\delta$ (where $\langle\dots\rangle_\delta$ indicates moments taken with the deviation $\delta f$ from $f_\mathrm{iso}$). In the last equation we introduced the notation $A^{\langle\mu\nu\rangle}\equiv \Delta^{\mu\nu}_{\alpha\beta} A^{\alpha\beta}$, with $\Delta^{\mu\nu}_{\alpha\beta} = \frac{1}{2}\bigl(\Delta^\mu_{\ \alpha} \Delta^\nu_{\ \beta} + \Delta^\nu_{\ \alpha} \Delta^\mu_{\ \beta}\bigr) - \frac{1}{3}\Delta^{\mu\nu}\Delta_{\alpha\beta}$, for the traceless and transverse (to $u^\mu$) part of a tensor $A^{\mu\nu}$. The shear stress tensor $\pi^{\mu\nu}\eq{T}^{\langle\mu\nu\rangle}$ has 5 independent components. Altogether, the deviation $\delta f$ has introduced 9 additional dissipative flow degrees of freedom. Their corresponding 9 equations of motion are controlled by microscopic physics and can be derived from approximate solutions of the Boltzmann equation \cite{Israel:1979wp,Denicol:2012cn}.

{\bf 3.} Anisotropic hydrodynamics \cite{Martinez:2010sc,Florkowski:2010cf} is obtained from (\ref{eq5}) by using in (\ref{eq3}) the spheroidally deformed local momentum distribution
\begin{equation}
\label{eq8}
  f(x,p) = f_\mathrm{RS}(x,p) \equiv 
  f_\mathrm{iso}\left(\frac{\sqrt{p_\mu\Xi^{\mu\nu}(x) p_\nu}-\tilde\mu(x)}{\Lambda(x)}\right),
\end{equation}
where $\Xi^{\mu\nu}(x) = u^\mu(x) u^\nu(x) +\xi(x) z^\mu(x) z^\nu(x)$, $z^\mu(x)$ being a unit vector in $z$ direction in the LRF. This distribution is characterized by 3 flow parameters $u^\mu(x)$ and three ``thermodynamic'' parameters: the ``transverse temperature'' $\Lambda(x)$, the effective chemical potential $\tilde{\mu}(x)$, and the momentum-anisotropy parameter $\xi(x)$. Inserting (\ref{eq8}) into (\ref{eq3}) yields the {\sc aHydro} decomposition
\begin{eqnarray} 
\label{eq9}
\!\!\!\!\!\!\!\!\!\!\!\!\!\!
  && j^\mu_\mathrm{RS} = n_\mathrm{RS} u^\mu, \quad 
  T^{\mu\nu}_\mathrm{RS} = e_\mathrm{RS} u^\mu u^\nu - \mathcal{P}_T \Delta^{\mu\nu} 
  + (\mathcal{P}_L-\mathcal{P}_T)z^\mu z^\nu,
\\\label{eq10}
\!\!\!\!\!\!\!\!\!\!\!\!\!\!
  &&n_\mathrm{RS} = \langle E\rangle_\mathrm{RS} = {\cal R}_0(\xi)\, n_\mathrm{iso}(\Lambda,\tilde\mu),
      \quad 
       e_\mathrm{RS} = \langle E^2\rangle_\mathrm{RS} = {\cal R}(\xi)\, e_\mathrm{iso}(\Lambda,\tilde\mu),
      \quad
      \mathcal{P}_{T,L} = \langle p^2_{T,L}\rangle_\mathrm{RS} = {\cal R}_{T,L}(\xi) \,
      \mathcal{P}_\mathrm{iso}(\Lambda,\tilde\mu).
\end{eqnarray}  
For massless systems, the local momentum anisotropy effects factor out via the $\mathcal{R}(\xi)$-functions, given in \cite{Martinez:2010sc}. The isotropic pressure is obtained from a locally isotropic EOS $\mathcal{P}_\mathrm{iso}(\Lambda,\tilde\mu)\eq\mathcal{P}_\mathrm{iso}(e_\mathrm{iso}(\Lambda,\tilde\mu),n_\mathrm{iso}(\Lambda,\tilde\mu))$. For massless noninteracting partons, $\mathcal{P}_\mathrm{iso}(\Lambda,\tilde\mu)=\frac{1}{3}e_\mathrm{iso}(\Lambda,\tilde\mu)$ independent of chemical composition. To compare with ideal and IS viscous hydrodynamics, we need to assign the locally anisotropic system an appropriate temperature $T(x)\eq{T}\bigl(\xi(x),\Lambda(x),\tilde{\mu}(x)\bigr)$ and chemical potential $\mu(x)\eq\mu\bigl(\xi(x),\Lambda(x),\tilde\mu(x)\bigr)$, thinking of $f_\mathrm{RS}(\xi,\Lambda)$ as an expansion around the locally isotropic distribution $f_\mathrm{iso}(T)$. For this we impose the generalized Landau matching conditions $e_\mathrm{RS}(\xi,\Lambda,\tilde\mu)\eq{e}_\mathrm{iso}(T,\mu)$ and $n_\mathrm{RS}(\xi,\Lambda,\tilde\mu)\eq{\cal R}_0(\xi)\,n_\mathrm{iso}(T,\mu)$. For example, using an exponential (Boltzmann) function for $f_\mathrm{iso}$ with $\mu=\tilde\mu=0$, one finds $T\eq\Lambda {\cal R}^{1/4}(\xi)$. With this matching we can write
\begin{eqnarray}
\label{eq11}
\!\!\!\!\!\!\!\!\!\!\!\!\!\!
   &&T^{\mu\nu}_\mathrm{RS} =  T^{\mu\nu}_\mathrm{id} 
   - (\Delta \mathcal{P}+\Pi_\mathrm{RS})\Delta^{\mu\nu} + \pi^{\mu\nu}_\mathrm{RS},
\\\label{eq12}
 \!\!\!\!\!\!\!\!\!\!\!\!\!\!  
  &&\Delta\mathcal{P} + \Pi_\mathrm{RS} = -\frac{1}{3}\int_p p_\alpha \Delta^{\alpha\beta} p_\beta
      (f_\mathrm{RS}-f_\mathrm{iso}) \qquad (= 0\ \mathrm{for}\ m=0),
\\\label{eq13}
\!\!\!\!\!\!\!\!\!\!\!\!\!\! 
  &&\pi^{\mu\nu}_\mathrm{RS} = \int_p p^{\langle\mu} p^{\nu\rangle}  (f_\mathrm{RS}{-}f_\mathrm{iso})
       = (\mathcal{P}_T{-}\mathcal{P}_L) \,\frac{x^\mu x^\nu+y^\mu y^\nu-2 z^\mu z^\nu}{3}.
\end{eqnarray}
We see that $\pi^{\mu\nu}_\mathrm{RS}$ has only one independent component, $\mathcal{P}_T{-}\mathcal{P}_L$, so {\sc aHydro} leaves 4 of the 5 components of $\pi^{\mu\nu}$ unaccounted for. For massless particles we have $(\mathcal{P}_T{-}\mathcal{P}_L)/\mathcal{P}_\mathrm{iso}(e)\eq{\cal R}_T(\xi){-}{\cal R}_L(\xi)$, so the equation of motion for $\pi^{\mu\nu}_\mathrm{RS}$ can be replaced by one for $\xi$. For $m\ne0$ we need an additional  ``anisotropic EOS'' for $(\Delta\mathcal{P}/\mathcal{P}_\mathrm{iso}){\,\equiv\,}(2\mathcal{P}_T{+}\mathcal{P}_L)/(3\mathcal{P}_\mathrm{iso}) - 1$, in order to separate $\Delta\mathcal{P}$ from the viscous bulk pressure $\Pi$.

{\bf 4.} Finally, {\sc vaHydro} \cite{Bazow:2013ifa} is obtained by generalizing the ansatz (\ref{eq8}) to include arbitrary (but small) corrections to the spheroidally deformed $f_\mathrm{RS}(x,p)$:
\begin{equation}
\label{eq14}
  f(x,p) = f_\mathrm{RS}(x,p) + \delta\tilde f(x,p) =
  f_\mathrm{iso}\left(\frac{\sqrt{p_\mu\Xi^{\mu\nu}(x) p_\nu}-\tilde\mu(x)}{\Lambda(x)}\right) 
  + \delta\tilde f(x,p) .
\end{equation}
The parameters $\Lambda$ and $\tilde{\mu}$ are Landau-matched as before, {\it i.e.} by requiring $\langle E\rangle_{\tilde\delta}\eq\langle E^2\rangle_{\tilde\delta}\eq0$; to fix the value of the defor\-ma\-tion parameter $\xi$ we demand that $\delta\tilde{f}$ does not contribute to the pressure anisotropy $\mathcal{P}_T{-}\mathcal{P}_L$, which requires $(x_\mu x_\nu{+} y_\mu y_\nu{-}2 z_\mu z_\nu)\langle p^{\langle\mu} p^{\nu\rangle}\rangle_{\tilde\delta}\eq0$. Then, upon inserting (\ref{eq14}) into (\ref{eq3}), we obtain the {\sc vaHydro} decomposition 
\begin{equation}
\label{eq15}
       j^\mu = j^\mu_\mathrm{RS} +\tilde{V}^\mu,\quad
       T^{\mu\nu}=T^{\mu\nu}_\mathrm{RS} - \tilde\Pi \Delta^{\mu\nu} + \tilde\pi^{\mu\nu},\quad
       \mathrm{with}\quad
       \tilde{V}^\mu = \bigl\langle p^{\langle\mu\rangle}\bigr\rangle_{\tilde\delta},\quad
       \tilde\Pi = -{\textstyle\frac{1}{3}} \bigl\langle p^{\langle\alpha\rangle} 
       p_{\langle\alpha\rangle}\bigr\rangle_{\tilde\delta},\quad
       \tilde{\pi}^{\mu\nu} = \bigl\langle p^{\langle\mu} p^{\nu\rangle}\bigr\rangle_{\tilde\delta},
\end{equation}
subject to the constraints $u_\mu \tilde\pi^{\mu\nu}\eq\tilde\pi^{\mu\nu} u_\nu\eq(x_\mu x_\nu{+}y_\mu y_\nu{-}2 z_\mu z_\nu)\tilde\pi^{\mu\nu}\eq\tilde\pi^\mu_\mu\eq0$. Clearly, the additional shear stress $\tilde\pi^{\mu\nu}$ arising from $\delta f$ has only 4 degrees of freedom. -- The strategy in {\sc vaHydro} is now to solve hydrodynamic equations for {\sc aHydro} (which treat $\mathcal{P}_T{-}\mathcal{P}_L$ nonperturbatively) with added viscous flows from $\delta\tilde f$, together with IS-like ``perturbative'' equations of motion for $\tilde\Pi,\,\tilde V^\mu$, and $\tilde\pi^{\mu\nu}$. The hydrodynamic equations are obtained by using the decomposition (\ref{eq15}) in the conservation laws (\ref{eq5}). The evolution equations for the dissipative flows $\tilde \Pi,\, \tilde V^\mu$, and $\tilde\pi^{\mu\nu}$ are derived by generalizing the procedure in \cite{Denicol:2012cn} to an expansion of the distribution function around the spheroidally deformed $f_\mathrm{RS}$ in (\ref{eq8}), using the 14-moment approximation. The equations are lengthy and found in \cite{Bazow:2013ifa}. We give their simplified form for (0+1)-d expansion in the next section. Especially at early times $\delta\tilde f$ is much smaller than $\delta f$, since the largest part of $\delta f$ is already accounted for by the momentum deformation in (\ref{eq8}). The inverse Reynolds number $\tilde{\mathrm{R}}_\pi^{-1}=\sqrt{\tilde\pi^{\mu\nu}\tilde\pi_{\mu\nu}}/{\cal P}_\mathrm{iso}$ associated with the residual shear stress $\tilde\pi^{\mu\nu}$ is therefore strongly reduced compared to that associated with $\pi^{\mu\nu}$, significantly improving the range of applicability of {\sc vaHydro} relative to standard second-order viscous hydrodynamics.

%%%%%%%%%%%%%%%%%%%%%%%%%%%%%%%%%%%%%%%%%%%%%%%%%%%%%%%
\vspace*{-3mm}
\section{Testing {\sc vaHydro} in (0+1)-dimensional expansion}
\label{sec3}
\begin{figure}[t!]
\begin{center}
\begin{minipage}{0.5\linewidth}
\includegraphics[width=\linewidth]{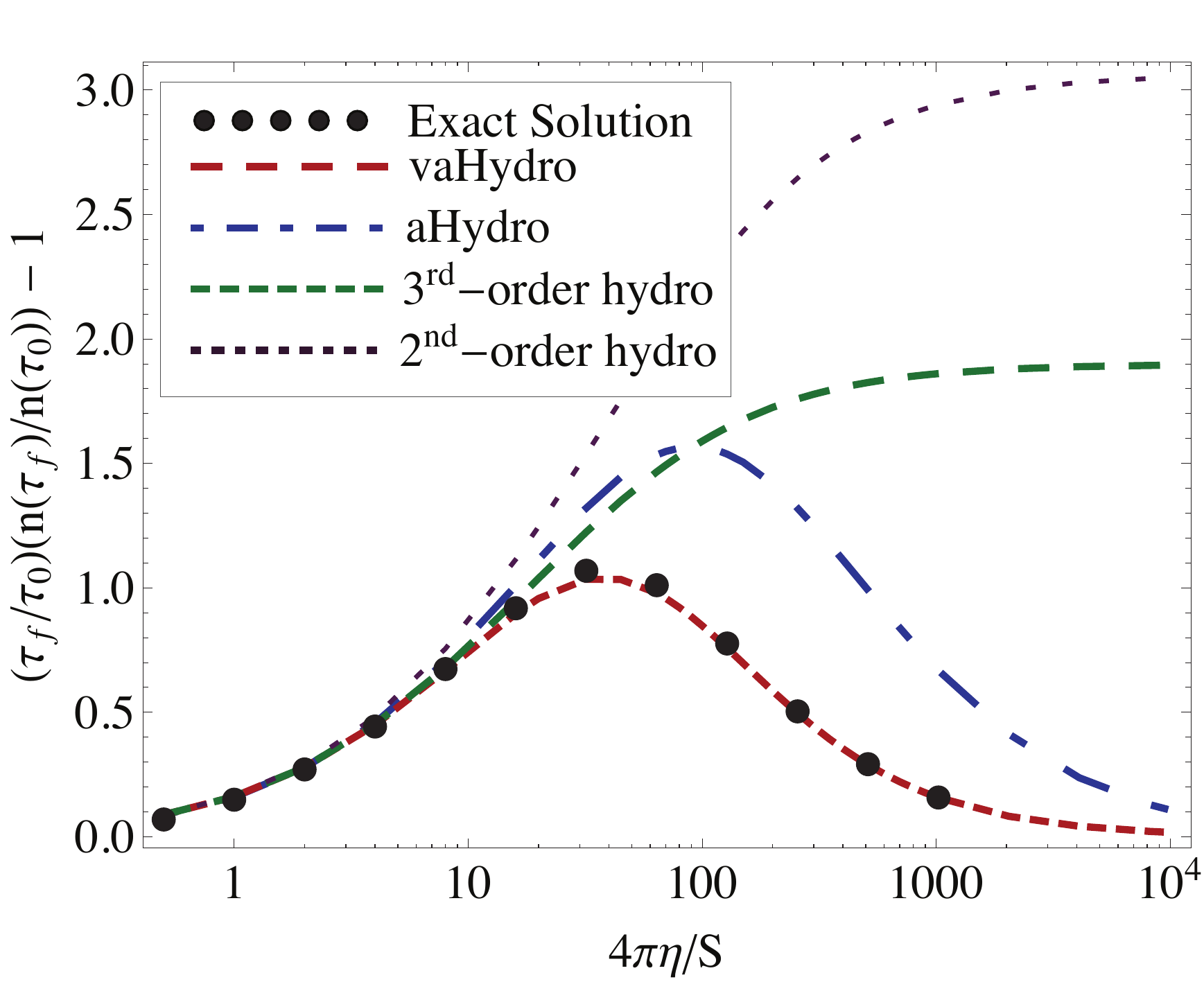}
\end{minipage}
\hspace*{0.03\linewidth}
\begin{minipage}{0.25\linewidth}
\caption{(Color online) The particle production measure $(\tau_f n(\tau_f))/(\tau_0 n(\tau_0)) -1$ as a function of $4\pi\eta/s$. The black points, red dashed line, blue dashed-dotted line, green dashed line, and purple dotted line correspond to the exact solution of the Boltzmann equation, {\sc vaHydro}, {\sc aHydro}, third-order viscous hydrodynamics \cite{Jaiswal:2013vta}, and second-order viscous hydrodynamics \cite{Denicol:2012cn}, respectively. The initial conditions are $T_0\eq600$\,MeV, $\xi_0\eq0$, and $\tilde\pi_0\eq0$ at $\tau_0\eq0.25$\,fm/$c$. The freeze-out temperature was taken to be $T_f\eq150$\,MeV.\newline
\label{F1}
}
\end{minipage}
\vspace*{-7mm}
\end{center}
\end{figure}

\vspace*{-1mm}
%%%%%%%%%%%%%%%%%%%%%%%%%%%%%%%%%%%%%%%%%%%%%%%%%%%%%%%
For (0+1)-d longitudinally boost-invariant expansion of a transversally homogeneous system, the Boltzmann equation can be solved exactly in RTA \cite{Florkowski:2013lya}, and the solution can be used to test the various macroscopic hydrodynamic approximation schemes. Setting homogeneous initial conditions in $r$ and space-time rapidity $\eta_s$ and zero transverse flow, $\tilde\pi^{\mu\nu}$ reduces to a single non-vanishing component $\tilde\pi$: $\tilde\pi^{\mu\nu}=\mathrm{diag}(0,-\tilde\pi/2,-\tilde\pi/2,\tilde\pi)$ at $z=0$. We use the factorization $n_\mathrm{RS}(\xi,\Lambda)\eq{\cal R}_0(\xi)\,n_\mathrm{iso}(\Lambda)$ etc. to get equations of motion for $\dot\xi,\, \dot\Lambda,\,\dot{\tilde\pi}$ \cite{Bazow:2013ifa}: 
\begin{eqnarray}
\!\!\!\!\!\!\!\!\!\!\!\!\!\!
&&\frac{\dot\xi}{1{+}\xi}-6\frac{\dot\Lambda}{\Lambda}=
\frac{2}{\tau}+\frac{2}{\tau_\mathrm{rel}}\left(1-\sqrt{1{+}\xi}\,{\cal R}^{3/4}(\xi)\right)\;,\quad
{\cal R}'(\xi)\, \dot\xi + 4 {\cal R}(\xi) \frac{\dot\Lambda}{\Lambda} =
- \Bigl({\cal R}(\xi) + {\textstyle\frac{1}{3}} {\cal R}_L(\xi)\Bigr) \frac{1}{\tau} 
+\frac{\tilde\pi}{e_\mathrm{iso}(\Lambda)\tau},
\\\nonumber
\!\!\!\!\!\!\!\!\!\!\!\!\!\!
&&\dot{\tilde\pi}=
-\frac{1}{\tau_\mathrm{rel}}\Bigl[\bigl({\cal R}(\xi){\,-\,}{\cal R}_{\rm L}(\xi)\bigr)P_\mathrm{iso}(\Lambda)+\tilde\pi\Bigr] -\lambda(\xi)\frac{\tilde\pi}{\tau}
   +12\biggl[
   \frac{\dot{\Lambda}}{3\Lambda}\Bigl({\cal R}_{\rm L}(\xi){\,-\,}{\cal R}(\xi)\Bigr)
  +\Bigl(\frac{1{+}\xi}{\tau}-\frac{\dot{\xi}}{2}\Bigr)
  \Bigl({\cal R}^{zzzz}_{-1}(\xi){\,-\,}\frac{1}{3}{\cal R}^{zz}_{1}(\xi)\Bigr)
  \biggr] P_\mathrm{iso}(\Lambda),
\end{eqnarray}
where $\lambda(\xi)$ and all the $\mathcal{R}$-functions can be found in \cite{Bazow:2013ifa}. $\tau_\mathrm{rel}$ and the ratio of shear viscosity $\eta$ to entropy density $s$, $\eta/s$, are related by $\tau_\mathrm{rel}\eq5\eta/(sT)\eq5\eta/(\mathcal{R}^{1/4}(\xi)s\Lambda)$. In \cite{Bazow:2013ifa} we solved these equations and compared with the exact solution, and also with the other hydrodynamic approximation schemes discussed above plus a 3rd-order viscous hydrodynamic approximation derived in \cite{Jaiswal:2013vta}. As an example, we show in Fig.~\ref{F1} the entropy production (measured by the increase in particle number $\tau n(\tau)$) between start and end of the dynamical evolution from an initial temperature of 600\,MeV to a final one of 150\,MeV. For this extreme (0+1)-d scenario, where the difference between longitudinal and transverse expansion rates is maximal, {\sc vaHydro} is seen to reproduce the exact solution almost perfectly, dramatically outperforming all other hydrodynamic approximations.  

%%%%%%%%%%%%%%%%%%%%%%%%%%%%%%%%%%%%%%%%%%%%%%%%%%%%%%%%
%%%%%%%%%%%%%%%%%%%%%%%%%%%%%%%%%%%%%%%%%%%%%%%%%%%%%%%%



\bigskip
\noindent
{\bf Acknowledgments: }
This work was supported in part by the U.S. Department of Energy, Office of Science, Office of Nuclear Physics, under Awards No. \rm{DE-SC0004286} and (within the framework of the JET Collaboration) \rm{DE-SC0004104}. Clarifying discussions with Gabriel Denicol and Dirk Rischke are gratefully acknowledged.

%% References


%\bibliographystyle{elsarticle-num}



\end{document}